# Composition gradients and their effects on superconductivity in Al-doped MgB$_2$

**A. J. Zambano, A. R. Moodenbaugh, and L. D. Cooley**

Materials Science Department, Brookhaven National Laboratory, 76 Cornell Ave., Upton NY 11973, USA

**Abstract**

A primary difference between pure MgB$_2$ and its alloyed forms is that the former is a line compound and, once formed, has the same composition everywhere, whereas the latter is a solid solution and requires diffusion to move alloying elements. Since defect energies are high, this opens up the possibility that alloying elements might not be distributed homogeneously, which could have important consequences for the observed superconducting properties. To address this issue, two sets of Mg$_{1-x}$Al$_x$B$_2$ samples, with $0 \le x < 0.5$, were prepared from elements using reaction temperatures and times at opposite extremes of those typically reported in the literature. Sample set *A* was given a reaction of 1 hour at 850 °C, which stopped just short of completion, while sample set *B* was reacted at temperatures as high as 1200 °C and thoroughly annealed for over 80 hours. The trace reactants remaining after reaction *A* indicated that Al is taken up more slowly than Mg, thereby making compositional gradients likely. Indeed, Williamson-Hall analyses of x-ray diffraction peaks showed that set *A* had higher crystalline strain than set *B* when $x > 0$ but *not* when $x = 0$. Since the presence of Al correlated with increased strain only for set *A*, it was concluded that reaction *A* produced substantial Al gradients across the individual grains while reaction *B* did not. Magnetization and heat capacity measurements indicated good bulk superconducting properties for all samples despite their structural differences, and consistent trends were observed when each sample set was considered alone. However, when both sets were considered together, their behavior was distinct when plotted vs. $x$ (e.g. two $T_c(x)$ curves), with trends for set *A* being shifted toward higher $x$ than for set *B*. On the other hand, all of the data merged (e.g. one $T_c(v)$ curve) when analyzed in terms of the unit cell volume $v$. Thus, while the first analysis might suggest that the different reactions produced different superconducting behavior, the second analysis, which captures the average Al content actually present inside the grains, shows that the samples have common behavior intrinsic to the addition of Al. Moreover, these analyses show that it is important to coordinate structural and property characterizations to remove artifacts of composition gradients and uncover the intrinsic trends. Because the standard characterizations of the superconducting properties above gave no clear indication that the two sample sets had different homogeneity, the structural information was vital to make a correct assessment of the effects of Al doping on superconductivity. Since many investigations have used reactions similar to reaction *A* and did not analyze data in terms of structural changes, previous results should be interpreted cautiously.

## 1. Introduction

Superconductivity in MgB$_2$ continues to generate a great amount of interest due to both its interesting physics and the real possibilities for applications. The achievement of long-length multifilamentary wires with high critical current densities by multiple groups[1,2,3,4,5] now makes it possible to begin to evaluate MgB$_2$ for niche applications, such as the generation of magnetic fields above the capability of copper-iron electromagnets without cooling by liquid cryogens.[6,7] However, the potential for broader impact still depends on the development of high-field and high-current forms of MgB$_2$ that can compete with other available materials and extend the field range of potential new applications at 20-30 K. Fortunately, there are strong indications that carbon-alloyed MgB$_2$ can support upper critical fields $H_{c2}$ above those that can be attained in Nb$_3$Sn[8] without expensive raw materials or complicated processing. Increasing $H_{c2}$ also improves the irreversibility field and extends the field range where critical current densities high enough for applications can be achieved.

Thin film studies[9,10,11] show clearly that the amount of electron scattering and $H_{c2}$ can be increased by alloying with carbon and retaining strain. High-field carbon-alloyed MgB$_2$ has also been reported for crystals[12,13] and in wires made with boron carbide,[14] nanoscale carbon,[15] or SiC (which is decomposed *in situ*).[16] Strain introduced by irradiation with neutrons[17], protons[18], or by ball-milling[19,20] likewise has a beneficial effect on the upper critical and irreversibility fields.

The development of conventional high-field superconductors over the past 50 years has relied on the fact that alloying with non-magnetic impurities does not cause significant reduction of the critical temperature $T_c$,[21] making it possible to shift the entire $H_{c2}(T)$ curve upward without compromising greatly the operating temperature margin. This does not appear to be the case for MgB$_2$, for which all attempts at chemical substitution, in particular Sc or Al for Mg[22,23,24,25,26,27] or C for B,[14,28,29,30,31] reduce $T_c$. Also, applying pressure[32] or introducing strain by irradiation[17,18] or mechanical damage[19] drives down $T_c$. Thus, the $H_{c2}(T)$ curve of MgB$_2$ does not shift uniformly upward upon alloying; instead it pivots around a point somewhat below $T_c$, significantly reducing the benefit for applications in the 20-30 K range. Understanding the reason for the loss of $T_c$ with alloying and investigating ways to offset this loss could improve the possibilities for applications at all temperatures.

Variations of $T_c$ as a function of strain and atomic substitution have been widely studied in intermetallic superconductors, in particular for compounds with the A15 crystal structure such as $Nb_3Sn$.[33] In contrast to metallic superconductors, covalent bonding in intermetallic compounds makes them much more sensitive to variations in the size and charge of substituted atoms, the displacement of atoms from their equilibrium sites, and to crystal lattice imperfections. These variations alter the density of electronic states (DOS) at the Fermi surface and the phonon spectrum, causing changes in the electron-phonon coupling and thereby $T_c$. The same is true for $MgB_2$; however, the two-band superconductivity mechanism[34,35] makes the response more complicated due to the combinations of intraband effects that are possible. In particular, hole states at the top of the important boron σ band lie close to the Fermi energy, and thus substitution of elements with extra electrons should have a strong effect on $T_c$.[36] Indeed, most successful alloying experiments so far obey this trend.[37] Moreover, interband scattering, insofar unique to $MgB_2$,[38] is an additional possibility for the reduction of $T_c$ that is additive to the other effects caused by changing the electron-phonon interaction. It is thus quite difficult to determine which of these possibilities is the primary cause of $T_c$ loss.

To make headway on this dilemma, the underlying complication of sample variability must be overcome. This is addressed in this article. As reviewed recently,[39] realities such as porosity make it difficult to identify which behavior is intrinsic to alloying and which is an artifact of sample quality. Mg volatility is difficult to suppress unless reactions are carried out in closed systems.[40] Since pure $MgB_2$ appears to be a line compound,[41] vacancies and other defects have high energies[42] and hinder the transport of atoms by diffusion. This may explain why nanoscale C and SiC precursors are required to achieve noticeable improvement in $H_{c2}$.[15,16] Reaction methods that produce homogeneous pure $MgB_2$ samples do not result in comparable homogeneity in doped samples,[26,28,31] as indicated by breadth in the superconducting transition and other characterizations. This suggests that uniform alloy composition throughout the entire sample is difficult to achieve unless very high temperatures and long reaction times are used.[14,28] In this case, since $T_c$ falls with increasing dopant content, pockets of weak superconductivity might be contained by a continuous region of strong superconductivity (or vice-versa), producing misleading trends as a function of the nominal dopant composition (i.e. as prepared in the starting element mixture). High defect energies also make the background level of lattice imperfections difficult to vary systematically by using near-equilibrium synthesis routes. This explains why many different bulk synthesis routes can be used to produce pure samples with good $T_c$ (between 37 and 39 K, depending on the boron purity[28]), whereas *in-situ* thin film synthesis and other non-equilibrium routes can produce $T_c$ as low as 20 K.[43] Other possibilities, such as coherent nanoprecipitates and their associated lattice strains,[44,45,46,47] oxygen interstitial phases,[48] dopant site disorder,[12] and superstructures,[49] are additional variables related to synthesis. Yet, *despite these realities, the superconducting properties in most experimental work are compared as a function of the amount of dopant material included in the sample preparation (nominal composition)*, which assumes (rather boldly) that all of the intended dopant is incorporated homogeneously in the $MgB_2$ phase.

In the experiments reported here, we show that Al-doped $MgB_2$ samples with different levels of overall homogeneity can be produced by different reactions, and these differences can be distinguished by routine analyses. We also show that, despite the differences in sample quality, intrinsic properties of the superconducting phase can be analyzed and common behavior as a function of Al concentration can be extracted. Trends plotted as a function of nominal composition are very different from those plotted as a function of the unit cell volume obtained by x-ray diffraction data. The analysis using unit cell volume provides a deeper insight, since it reflects the average composition actually present in the grains. Without considering these differences, it would be very easy to misinterpret the changes that occur upon alloying, and we discuss one example. Although the experiments are restricted to Al doping, our methodology and conclusions can be more generally applied to other dopants of $MgB_2$.

## 2. Experiment

### A. Synthesis and structural characterization

Powders of Mg (Alfa Aesar, −325 mesh, purity 99.8 %) and crystalline B (Alfa Aesar, −325 mesh, 99 % purity) and granular Al (Balzers, ~100 μm, purity 99.99%) were mixed in stoichiometric ratios to form a series of alloys $Mg_{1-x}Al_xB_2$, with $0 \leq x \leq 0.45$. The mixtures were pressed at ~7 kbar into ~10 mm tall by 9 mm diameter pellets, wrapped in Ta foils and placed in closed stainless steel crucibles under a 1 bar Ar atmosphere. The compressed powders were then processed following two different temperature-time regimens: (*A*) heating up at a rate of 5 °C/min up to 850 °C, with a 1-hour soak followed by furnace cooling over ~2 hours; (*B*) heating up at a rate of 12 °C/min up to 1200 °C with a 1-hour soak, followed by slow cooling down to 700 °C at 0.1 °C/min with a 5 or 60 hour soak (60 h in case of $0.25 \leq x \leq 0.45$ samples), followed by furnace cooling over ~2 hours. Note that these reactions span the range of temperature and time combinations reported in the literature, and that reaction *A* is similar to that often used to survey the effects of Al doping.

Microstructures surveyed by scanning and transmission electron microscopy (SEM and TEM) Demonstrated that the two sample sets showed dissimilar morphology. Samples made by reaction *A* produced mostly thin plate-like crystallites, with only a few that had parallelepiped shapes, as shown in Fig. 1(a). By contrast, reaction *B* gave well-defined parallelepiped-like crystallites with rounded corners, shown in Fig. 1(b). These differences did not seem to depend on the presence of Al. Because the actual grain dimensions are not clear in the SEM images in Fig. 1a, specimens of different samples were crushed and suspended on a graphite mesh and then viewed by TEM. The typical grain size when viewed in this manner was ~1.3 μm and as large as 10 μm for reaction *B*. For reaction A, it was still very difficult to establish a characteristic grain size due to the large aspect ratio. Roughly, grain sizes were less than 0.2 μm thick and smaller than 1 μm in diameter.

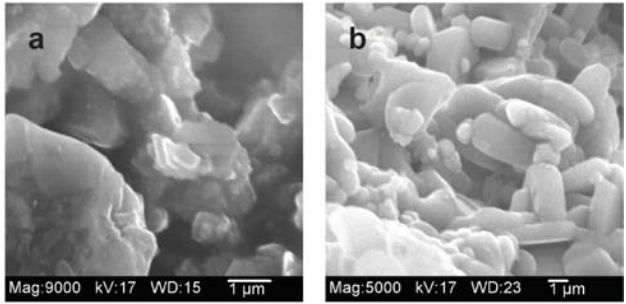

Fig. 1. Scanning electron microscopy images of fractured surfaces of samples with $x = 0.15$ from reaction $A$ (image (a) at left) and reaction $B$ (image (b) at right). Note the differences in the scale bars on the images.

Energy-dispersive x-ray spectroscopy (EDS) measurements taken in the SEM showed that the Al content in the $MgB_2$ grains of sample set $B$ fell close to the intended value, as shown in Fig. 2. Each data point represents an average composition for typically 10 different measurements on different grains. Similar measurements were not possible for sample set A because of the small grain size and the presence of impurity phases.

Powder x-ray diffraction patterns (XRD) were taken using a commercial diffractometer (Phillips 3100) using crushed powders from each sample and Si powder as a reference. Full diffraction patterns are shown in Fig. 3. These XRD patterns exhibit dominant peaks due to $Mg_{1-x}Al_xB_2$. The presence of Mg-Al phases is seen for most samples of set $A$ in Fig. 3(a), which indicates the solidification of a Mg-Al melt during cooling that is leftover from the incomplete reaction. There are also trace levels of MgO and $MgB_4$. The various Mg-Al phases, $Al_{12}Mg_{17}$, $Al_3Mg_2$, and (Al) with dissolved Mg, are noteworthy, because each has an Al content that is significantly higher than the corresponding nominal value of $x$ for the particular reaction. This trend indicates a slower uptake of Al than Mg by boron during reactions. Since this would produce an Al-rich crust on the outside of $(Mg,Al)B_2$ grains as Mg is consumed, significant compositional gradients are likely to be present in sample set $A$. Also, the amount of Al actually introduced into the $(Mg,Al)B_2$ grains on average is less than the intended amount based on the observed composition of the leftover reactants. On the other hand, sample set $B$ appears to be fully reacted in Fig. 3(b), showing peaks of only the alloyed $MgB_2$ phase and some trace peaks of $MgB_4$ and MgO. Additional peaks for $x > 0.15$ for sample set $B$ also appear which could not be traced to any identified product of elements present in the reaction or its container.

The x-ray diffraction data overall show systematic changes of the peak position and breadth, indicating variations of both crystal lattice parameters and in the crystalline perfection as a function of Al content. Previous work has shown that Al alloying produces a significant reduction of the $c$ axis and a smaller reduction of the $a$ axis.[22] This variation is also observed in our experimental data. Fig. 4 shows the relative change of the unit cell volume $v(x) = V(x)/V_0$ as a function of $x$, where $V_o = V(0)$ is the unit cell volume for pure $MgB_2$, and $V(x)$ is the unit cell volume obtained from the XRD data for Al-doped samples. Notice that $v(x)$ for samples in set $B$ generally follow the dashed line describing Vegard's law (VL), which is a simple linear mixing rule between $MgB_2$ and $AlB_2$. This suggests that the long annealing used in reaction $B$ produces samples with composition close to the intended one. By contrast, the data for sample set $A$ lie significantly above the VL line, with the displacement becoming stronger for higher $x$. This indicates that the Al content actually present within the $(Mg,Al)B_2$ grains is below the nominal composition for set $A$.

The analyses above suggest that strain due to crystalline imperfections and compositional gradients could be present in the samples. To explore this possibility, it is necessary to examine the breadth of the XRD peaks. The diffractometer that was used has a Cu radiation source, with two primary wavelengths $\lambda$ from the transitions $K\alpha_1 = 1.5405$ Å and $K\alpha_2 = 1.54434$ Å. This appears in the diffraction spectrum as a superposition of two Bragg peaks for each set of Miller indices $(hkl)$. Thus, in order to separate the $K\alpha_1$ and $K\alpha_2$ peaks for each reflection, two pseudo-Voigt functions (a Lorentzian function with weight $(1-\eta)$ plus a Gaussian function with weight $\eta$, with fitting parameter $0 \leq \eta \leq 1$) were combined to fit those $MgB_2$ reflections that emerged mostly separated from the peaks of impurity phases. The centroid of peak positions $(2\theta_{peak})$ and their width at half maximum $(\Delta_{peak})$, and $\eta$ could subsequently be determined. An example of the fit quality is displayed in Fig. 5. After the deconvolution of the diffraction peak, the pure full-width at half-maximum intensity of the peak $W$ was obtained from

$$W = \sqrt{(\Delta_{peak})^2 - (\Delta_{inst})^2}. \quad (1)$$

Here, $\Delta_{inst}$ is the instrument resolution determined by the x-ray source, instrument geometry, misalignment, etc., as estimated by measuring the broadening of the Bragg peaks as a function of angle $2\theta$ for a NIST standard.

The Williamson-Hall (W-H) method[50] was then used to extract information from the peak shapes and their variation with $2\theta$. In the W-H method, grain size and strain effects are separated by plotting the breadth of the reciprocal lattice points, $\beta^* = W\cos\theta/\lambda$, against their distance from the origin, $d^* = 2\sin\theta/\lambda$. These two quantities have the following relationship:

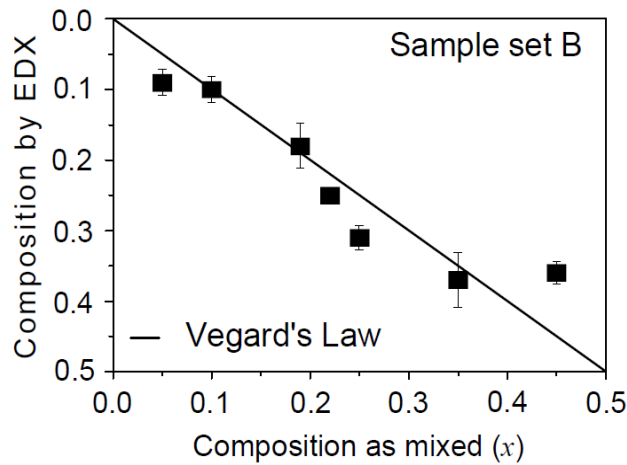

Fig. 2. Composition of individual grains for samples made by reaction $B$, as measured by energy-dispersive x-ray spectroscopy in an SEM, is plotted as a function of the nominal composition. Vegard's law is also plotted as a line.

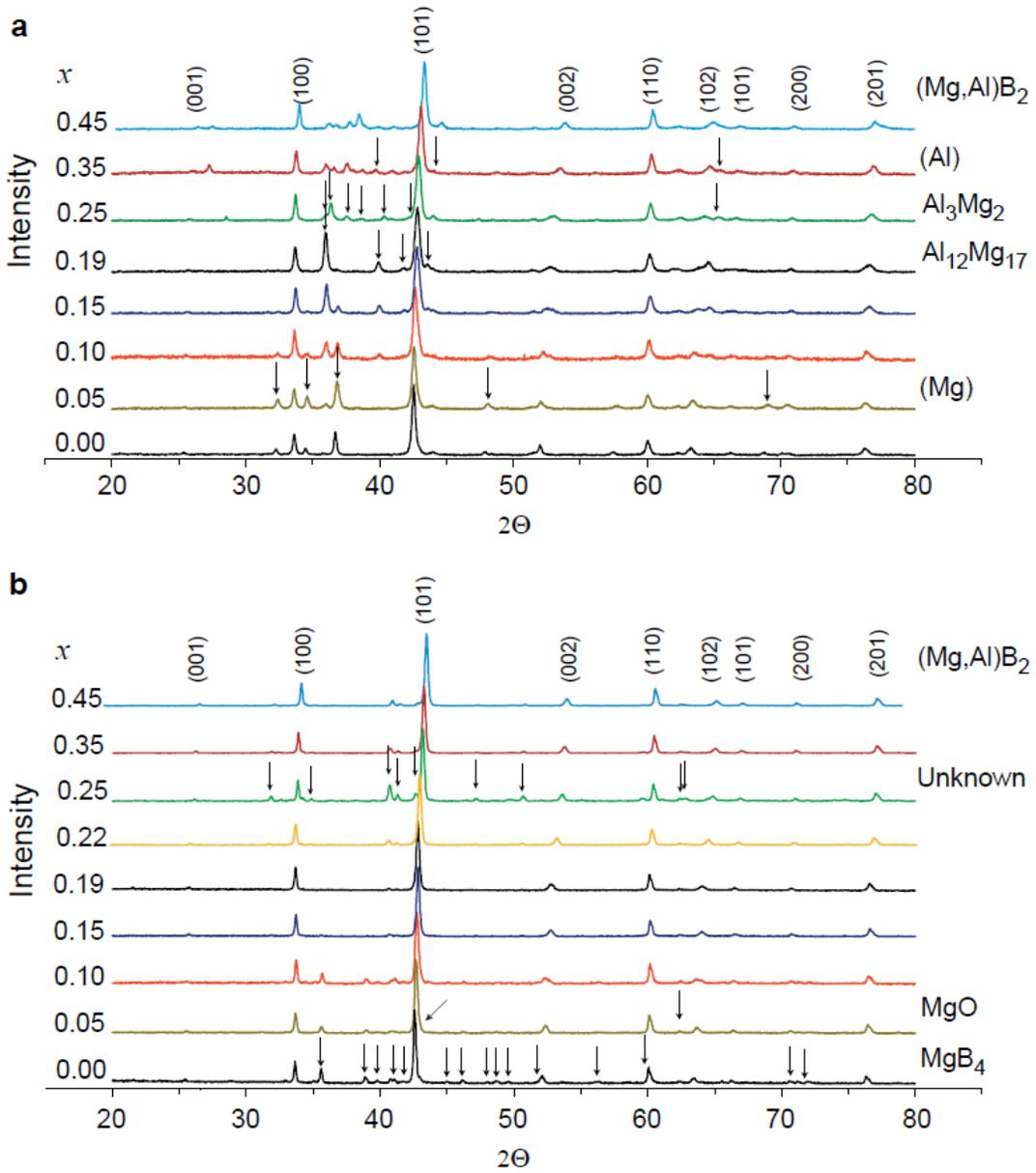

Fig. 3. X-ray diffraction data is shown for samples made by reaction *A* (top series of curves, (a)) and by reaction *B* (bottom series of curves, (b)). The nominal composition is labeled at the left of each spectra, and all spectra have been offset to provide clarity. Peak positions corresponding to the main $Mg_{1-x}Al_xB_2$ phase are labeled on the top curves of (a) and (b). Impurity phases are indicated by additional peaks appearing on many curves. These peaks are labeled on the curve where they appear most strongly, with the corresponding phase identified at the right when possible.

$$\beta^* = \frac{1}{D} + \frac{\varepsilon}{2}d^* \qquad (2)$$

where $D$ is an effective crystallite size and $\varepsilon$ is an effective strain. Fig. 6 shows the dependence of $\beta^*$ on $d^*$ for four Bragg reflections for samples from both heat treatment groups. Fig. 6(a) shows data for the undoped samples for both reactions, while Figs. 6(b) and 6(c) show data for doped samples from set *A* and *B* respectively. Since the addition of Al changes the *c* axis more strongly than the *a* axis, the (*hk*1) peaks show higher broadening than the (*hk*0) peaks do. This difference and the lack of other clear peaks in the x-ray diffraction patterns limit the interpretation of these plots. Nonetheless, it is still clear that in all plots the data points generally lie along a line with nonzero slope and intercept, which indicates that both the grain size and imperfections and distortions of the crystal structure contribute to the strain and the broadening of the XRD peaks.

Fig. 6(b) is clearly different from either Fig. 6(a) or Fig. 6(c). The higher slope in Fig. 6(b) than in Fig. 6(a) indicates that the presence of Al is related to the higher strain for samples that are just formed. However, the similarity of the slopes in Fig. 6(a) and Fig. 6(c) indicates that Al itself is not a significant source of strain when the samples are annealed well. It can be concluded that the main reason for the different slope in Fig. 6(b) is the inhomogeneous distribution of Al. These trends are summarized in Fig. 6(d), which shows the strain obtained from weighted least-squares fits of $\beta^*(d^*)$ for each sample as a function of *x*. There appears to be negligible dependence of $\varepsilon$ on *x* for sample set *B*, for which the strain hovers around the 0.4% value characteristic of the two pure samples. By contrast, $\varepsilon$ rises sharply for small $x$ and then drops gradually with increasing $x$ for sample set $A$. Except for the anomalous data point at $x = 0.25$, all of the data for doped samples in set $A$ have strains of ~0.8%, double that of the undoped sample. Thus, the distortions caused by the Al composition inhomogeneity are significant, being much larger than the pre-compression found in copper-sheathed Nb$_3$Sn wires, for instance.

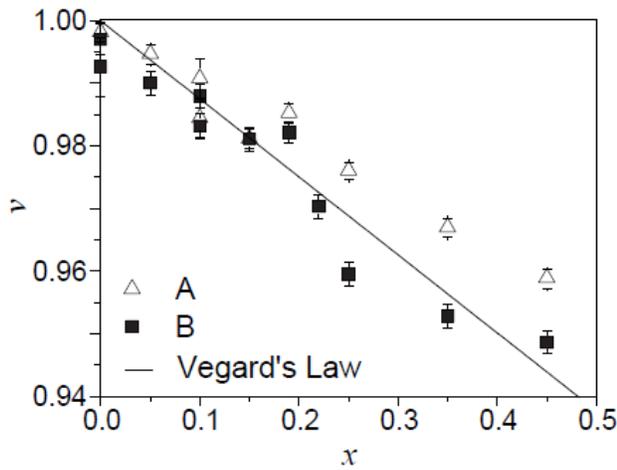

Fig. 4. The normalized unit cell volume is plotted as a function of the nominal Al composition for samples made by both reactions. The line represents Vegard's law, using the lattice parameters of MgB$_2$ and AlB$_2$.

## B. Superconducting property characterization

A commercial SQUID magnetometer (Quantum Design MPMS) was used to measure the dc magnetization at 1 mT applied field to determine $T_c$. The transition to the normal state upon warming after cooling in zero field was rather sharp for all samples, as indicated in Figs. 7(a) and 7(b), with the width $\Delta T_c$ between 90% and 10% of the magnetization at 5 K being 2 to 5 K for sample set A and 0.3 to 2.5 K for sample set B. The magnetization at 5 K was saturated for all samples except those for which $T_c$ was below 10 K, where lower temperature was needed to obtain a similar saturated state. The critical temperature is defined by the temperature at which the steepest slope in the middle of the transition extrapolates to zero magnetization. A slight difference in the breadth of the inductive transitions for the sample sets is apparent, with those for set $A$ being somewhat broader than the corresponding samples from set $B$ with the same nominal composition. There does not seem to be any apparent systematic degradation of the quality of the inductive transition with increasing $x$ for sample set $B$, while the transitions for $x > 0.20$ for set $A$ are somewhat broader than those for $x < 0.20$ in the same set. All of the samples reached magnetic moments that represented full shielding, determined from the applied field and the sample shape, mass, volume and porosity (determined by pyncnometry). Thus, shielding currents that determine the magnetization signal appear to flow uniformly within the bulk.

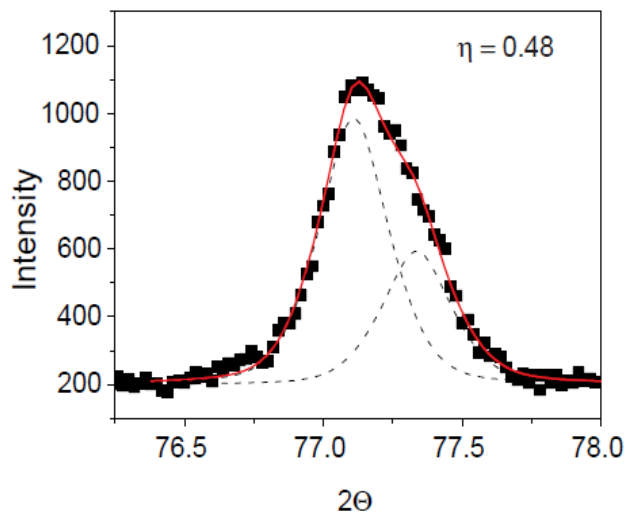

Fig. 5. An example of the fitting procedure used to determine the widths of the x-ray diffraction peaks. The data points are from the (002) peak for $x = 0.15$, sample set $A$. The solid red curve is a fit to the data, which is a convolution of the two pseudo-Voigt functions shown by the dashed curves with the weighting indicated in the top right corner. The width at half-maximum of the dashed curves was then used to define $W$.

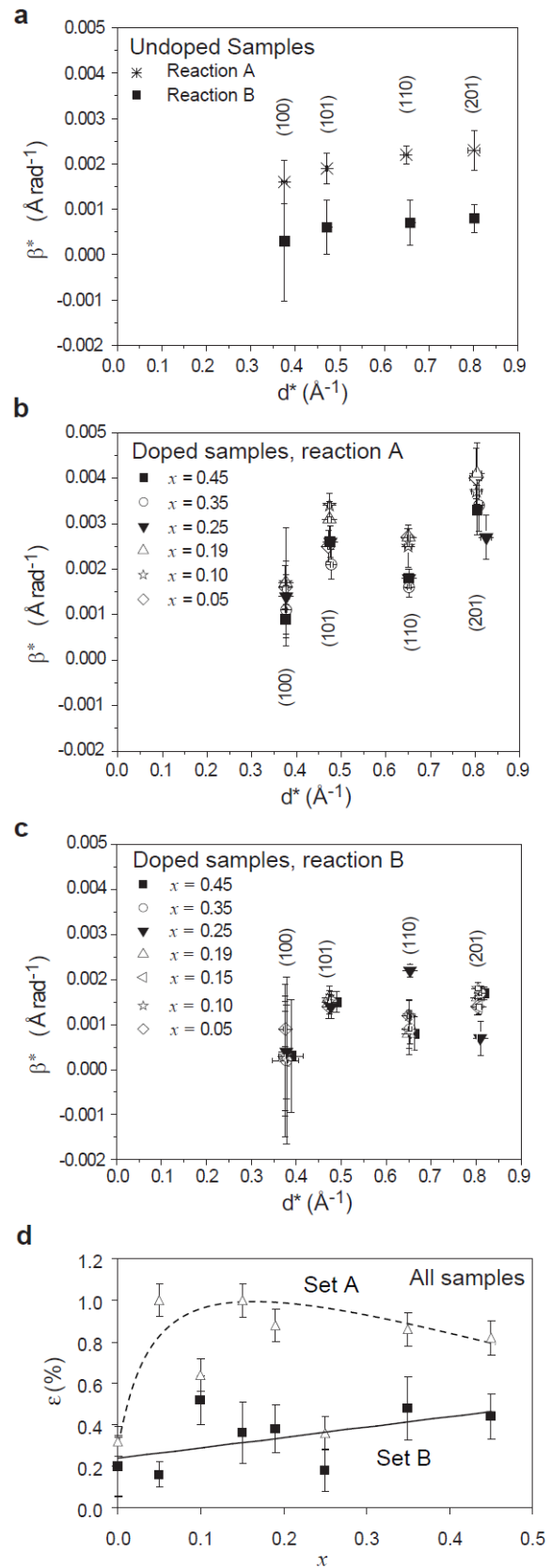

Fig. 6. Williamson-Hall plots, as described in the text, are shown for pure samples (plot (a)), Al-doped samples made by reaction $A$ (plot (b)), and Al-doped samples made by reaction $B$ (plot (c)). The different samples and their corresponding diffraction peaks are indicated on the plots. Plot (d) summarizes these data, where strains are calculated with the inclusion of the uncertainty bars in plots (a)-(c).

The dependence of $T_c$ on $x$ is stronger for sample set $B$ than for set $A$, as shown in Fig. 7(c), where there are 2 distinct curves, one for each sample set. However, as noted earlier, the nominal composition does not reflect the true Al content within the superconducting grains for sample set $A$. An improved representation of the dependence of $T_c$ on the average Al content is shown in Fig. 7(d), which compares the relationship between $T_c$ and $v$. Here, data for both sample sets collapse onto a single curve, suggesting that this plot captures the intrinsic behavior of the superconducting state.

Measurements of the heat capacity $C_p(H,T)$ were also made between 2 to 50 K using a commercial Quantum Design system, which implements a relaxation method described in detail elsewhere.[51] An example of the raw data is shown in Fig. 8(a). Here $H$ is the applied magnetic field and $T$ is temperature. Procedures to examine 2-band superconductivity in $Mg_{1-x}Al_xB_2$ follow: To good approximation, the heat capacity of the normal state $C_n$ can be obtained over a wide temperature range by the $C_p(T)$ curve measured at $\mu_oH_o \sim 8.75$ T.[52] The intercept of $C_n/T$ by extrapolating $T \to 0$ then provides the Sommerfeld coefficient $\gamma_n$ corresponding to the linear term of the electronic specific heat. The resulting curve with the normal state subtracted, $C_P(0,T) - C_P(H_o,T)$, expresses the electronic specific heat difference between the superconducting and normal states, $C_{es}(T) - C_{en}(T)$, including the significant jump at $T_c$. We used the $\alpha$ model[53] to fit the subtracted data, following procedures described in detail by Putti et al.[25] One example of the subtracted data and its fit is shown in the inset of Fig. 8(a). The fitting routine uses four parameters, three of which are independent: $\gamma_\sigma/\gamma_n$, $\gamma_\pi/\gamma_n$, $\Delta_\pi/(k_BT_c)$, and $\Delta_\sigma/(k_BT_c)$. These parameters express the energy gaps, $\Delta_\sigma$ and $\Delta_\pi$, and the Sommerfeld coefficients, $\gamma_\sigma$ and $\gamma_\pi$, in terms of each of the superconducting bands.

Fig. 8(b) shows the behavior of the energy gaps as functions of $x$. The $\pi$ and $\sigma$ gaps for samples made by reaction $A$ decrease smoothly toward 0 at $x \approx 0.45$, which is consistent with the nominal composition where $T_c$ extrapolates to 0 in Fig. 7(c). Samples made by reaction $B$ display a remarkably different behavior, with both gaps apparently falling abruptly at $x \sim 0.25$. However, this critical composition is also consistent with the $T_c(x)$ curve for reaction $B$ samples in Fig. 7(c).

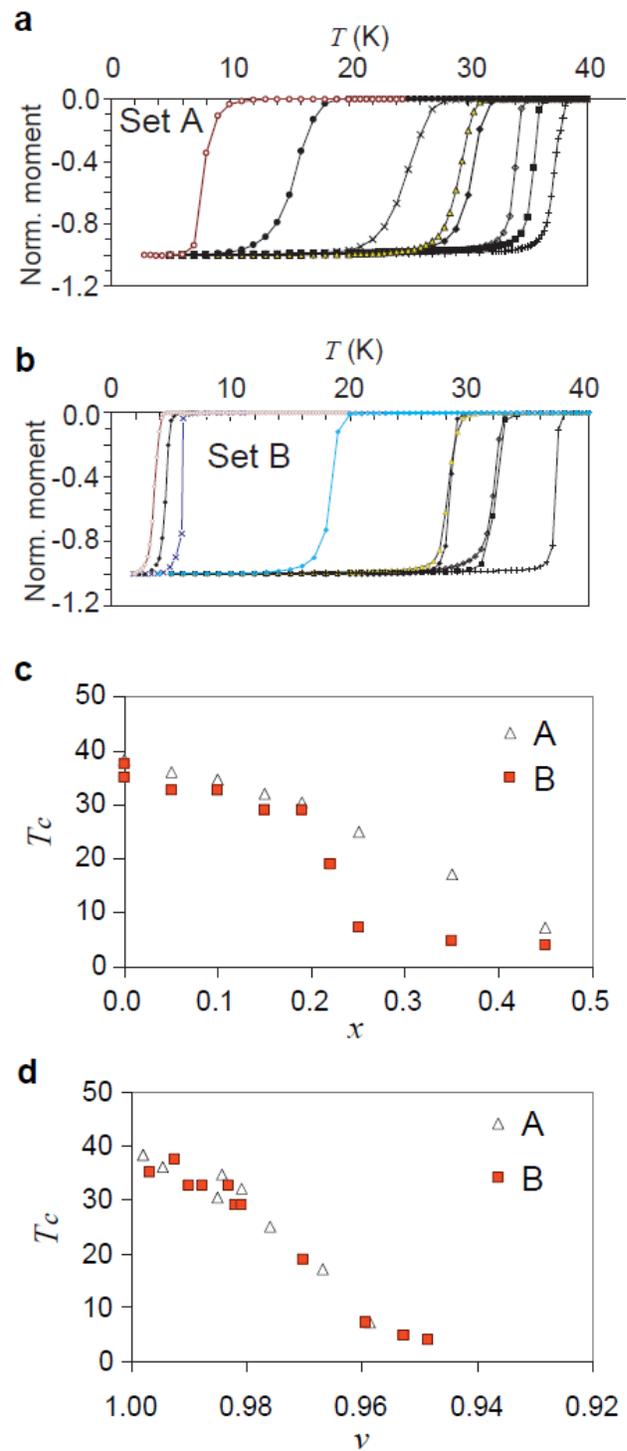

Fig. 7. Magnetization transitions measured upon warming in a 1 mT field after cooling in zero field are shown in plot (a) for samples made by reaction $A$, and in plot (b) for samples made by reaction $B$. For both plots, the transitions from left to right correspond to $x = 0.45, 0.35, 0.25, 0.22$ (plot (b) only) 0.19, 0.15, 0.10, 0.05, and 0.00. The data are normalized to the magnetization at low temperature. Plots (c) and (d) show the variation of $T_c$ with composition and with unit cell volume, respectively, for both sample sets.

Hence, assuming for the moment that the nominal composition represents the actual composition in the grains (this is not true for samples made by reaction $A$), magnetometry and heat capacity measurements appear to be self-consistent within a given sample set. There also appears to be significant differences between sets caused by the different reactions.

Figs. 8(c) and 8(d) show, however, that universal behavior emerges when the energy gaps are plotted as functions of $v$ and $T_c$. In these cases, a unique merging point for the gaps is found, corresponding to $T_c \approx 7$ K and $v \approx 0.96$. These merging points can further be identified with a unique point on Fig. 7(d) where the curvature of $T_c(v)$ changes. While the data in Figs. 8(c) and 8(d) do not coincide as tightly as they do in Fig. 7(d), the emergence of common trends *despite the structural differences between sample sets* indicates that this behavior is an intrinsic property of the Al doping. The unit cell volume at which the gaps appear to have equal values corresponds by Vegard's law to $x \approx 0.33$. In addition, notice that a gap or gaps persist (albeit with large uncertainty) for smaller unit cell volume and $T_c$ lower than the merging point, extrapolating out to $v \sim 0.935$ ($x \sim 0.52$) where superconductivity disappears completely. Again, this is consistent with the trend in $T_c(v)$ data in Fig. 7(d).

## 3. Discussion

An initial surprise shortly after the discovery of superconductivity in MgB$_2$ was the relative ease by which samples of apparently good quality could be made by reaction of Mg vapor with solid B. Reactions very similar to reaction *A* in this paper produced samples that exhibited full Meissner effects, sharp inductive transitions, reasonably high critical current densities, and an absence of weak links between grain boundaries.[54,55] The fact that pure MgB$_2$ does not deviate from stoichiometry causes uniform properties to appear wherever MgB$_2$ is formed, and the pervasiveness of Mg vapor (or liquid in sealed systems under modest pressure[40]) produces reactions throughout the volume of pressed and sintered powder samples. Unfortunately, such samples are not amenable to electron microscopy, because their nanoscale grain size and high porosity make microscopy sample preparation difficult. A goal of reaction *B* was initially to produce much larger grain size for structural and microchemical studies, following an earlier study by Badr and Ng.[56]

The ease of synthesis of pure MgB$_2$ does not immediately carry over to the fabrication of doped samples. With the possible exception of carbon, all doped forms of MgB$_2$ must be synthesized by bulk diffusion, which should be slow in the 800-1000°C temperature range due to the high energies for the formation of vacancies.[42] Much higher temperatures and longer times are required to achieve homogeneous samples, such as reaction *B* in the present paper. This fact was explicitly pointed out by Ribeiro et al.[28] Nonetheless, it is often the case that reactions similar to those used for synthesis of the pure binary compound, i.e. like reaction *A*, are used to make doped samples.

In the case of carbon doping, it is possible to begin with an atomic mixture of boron and carbon and overcome difficulties of bulk diffusion of C through B or MgB$_2$. There has been a variety of synthesis experiments using boron carbide compounds,[14,28,29] but even in that case, achieving properties as good as those for reactions of just Mg with B is difficult. Since B$_4$C itself retains the same crystal structure over a wide range of carbon content, roughly 7 to 20% C, and entropy may play a significant role in preventing homogeneous car-bon content from occurring,[57] diffusion might still be required to obtain a homogeneous distribution of carbon atoms.

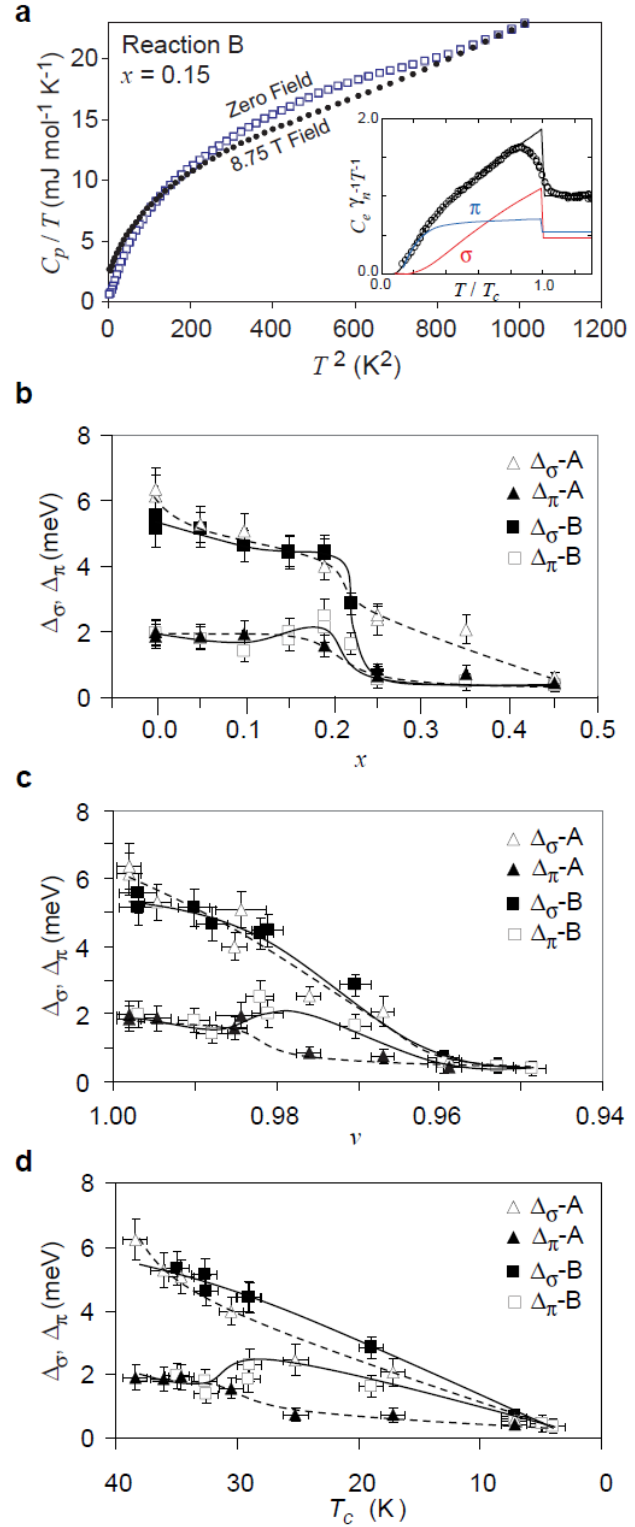

Fig. 8. An example of raw heat capacity data is shown in plot (a), for $x = 0.15$ in sample set *B*. The inset of plot (a) shows the reduced data after subtracting the lattice contribution. The black line is a fit to the reduced data using the $\alpha$ model, which is a sum of the red and blue curves giving the $\sigma$ and $\pi$ band contributions as indicated. Plots (b)-(d) show the magnitude of the superconducting gaps for each band derived from the $\alpha$ model, as functions of composition, unit cell volume, and $T_c$ respectively. Dashed curves are guides to the eye for $\Delta_\sigma$-A and $\Delta_\pi$-A, and solid curves are guides to the eye for $\Delta_\sigma$-B and $\Delta_\pi$-B.

At the start of the reactions among Mg, Al, and B considered here, there is already deviation from the intended composition of the $Mg_{1-x}Al_xB_2$ product, as indicated by the presence of $Al_{12}Mg_{17}$, $Al_3Mg_2$, and the (Al) solid solution phases after reaction *A*. Since the reactions were carried out in a container sealed at 1 bar Ar pressure at 300 K, the Ideal Gas Law and the thermodynamic calculations in ref. 40 indicate that the metal reactants are liquid and not vapor up to ~1300 °C, so these phases are probably solidified from an Al-rich melt. Given this imbalance at the start of the diffusion reaction, the quality of samples obtained at the end of the reaction depends on the time and temperature, and thus the completeness of the diffusion reaction. Since the Al penetrates more slowly from the outside, the most plausible scenario is that individual grains start out with a gradient of Al composition that is richer in Al on the outside. After some time at high temperature, the samples become more homogeneous.

The sample inhomogeneity has two important aspects for the interpretation of superconducting property measurements. First is the fact that $x$ does not accurately represent the overall Al content of the superconducting phase for sample set *A*. The solidified Al-rich melt left behind after reaction *A* produces an offset between the actual composition and $x$. The ~0.8% strain due to Al composition gradients further indicates that the Al concentration can vary by ±0.05 around the offset value, based on the ~12% difference in unit cell volumes between $MgB_2$ and $AlB_2$. Since the inductive transition widths are small, typically less than 5% of $T_c$, the offset and uncertainty in $x$ are the chief reasons why the data sets in Fig. 7(c) and in Fig. 8(b) do not coincide. Here, the x-ray diffraction data is very useful because the unit cell volume can be related to the average Al content in the superconducting phase via Vegard's law. This corrects for the inhomogeneity, yielding $T_c(v)$ data for both sample sets that fall onto a universal curve (Fig. 7(d)) and heat capacity data that overlap when plotted in terms of $v$ or $T_c$ (Figs. 8(c) and 8(d)). Thus, it is possible to overcome the problems introduced by gradients by combining structural and property measurements appropriately. This is the most important result of this paper.

The second aspect is the more subtle effect of gradients on the superconducting properties themselves. Recent studies of $Nb_3Sn$ superconductors[58,59] argued that resistive and inductive $T_c$ measurements probe systems of superconducting pathways that incorporate different components of the microstructure. Disparities can arise when regions with higher $T_c$ form a continuous network, because they can shield the regions with lower $T_c$. On the other hand, when regions with lower $T_c$ form a network, they cannot shield the regions with higher $T_c$ and the whole sample becomes transparent to magnetic probes. In the latter case, it is possible to compare explicitly the width of the inductive transition to the intrinsic variation of superconducting properties with composition.

In the present work, the exterior portions of grains are rich in Al, and therefore these regions form current pathways with lower $T_c$ than the bulk average. This makes the sample transparent to magnetic probes. However, it is more difficult to make a direct connection between the inductive measurements and the composition gradients because it is always possible for current loops to surround pores. Nonetheless, the inductive transitions in Fig. 7(a) are broader than the transitions in Fig. 7(b) because the gradients are probed by the induced currents. In this case, the way in which $T_c$ is defined incorporates the effects of Al inhomogeneity to varying degrees. A common definition of $T_c$ is the temperature corresponding to the onset of shielding, which emphasizes the Mg-rich portions of the samples and overlooks the Al-rich portions. Another definition is the temperature at which 50% shielding occurs, which averages the gradients more effectively but underestimates the temperature at which the resistance falls to zero. The definition in this paper is a compromise between these limits.

It is also important to notice that, had the additional information from XRD not been obtained for these samples, the trends in Figs. 7(c), and 8(b) could be interpreted differently. This implies that caution should be exercised when interpreting previous work, since many published studies of Al-doped $MgB_2$ used reactions similar to the present reaction *A* and reported findings in terms of the nominal composition. For example, the variations of $\Delta_\sigma$ and $\Delta_\pi$ with $x$ for sample set *A* (Fig. 8(b)) indicate a rather slow loss of superconductivity and no merging of the gaps. Slightly faster rates of decline but overall similar behavior were reported by Putti *et al.*,[25] who used a reaction between those in this paper, 1000 °C for 24 h. These data contrast strongly with $\Delta_\sigma(x)$ and $\Delta_\pi(x)$ for sample set *B*, which exhibits a steep drop at $x \sim 0.25$ and signs that $\Delta_\pi$ is enhanced at the expense of $\Delta_\sigma$, perhaps merging near the drop. These changes might suggest that the nature of 2-band superconductivity can be altered by changing the reaction temperature at given composition. On the other hand, when these data are plotted as a function of $v$ (Fig. 8(c)), it is seen that the sample sets are really not very different. A single trend is seen for both $\Delta_\sigma(v)$ and $\Delta_\pi(v)$, which indicates that this behavior is intrinsic to the addition of Al as we argue in another work.[60] Signs of enhancement of $\Delta_\pi$ at the expense of $\Delta_\sigma$ remain (this was predicted in ref. 61), suggesting that this is a real effect that deserves more attention.

It is interesting that the strain is highest for sample set *A* in the composition range where superstructures have been found. The reduced entropy associated with superstructures should slow the thermodynamic driving potential for the intermixture of Al on the Mg sublattice, which could contribute to the slower Al uptake. Although the XRD does not show any strong indication of superstructures for samples made by reaction *B*, several grains with superlattice diffraction spots were indicated by TEM, suggesting that the superstructure may not be completely overcome at higher temperature. Karpinski et al. recently reported problems with phase separation even at temperatures above 1500 °C in the growth of Al-doped $MgB_2$ crystals,[26] suggesting that the mixing entropy is still not high enough to obtain a single homogeneous composition.

## 4. Conclusions

Two different sets of samples of $Mg_{1-x}Al_xB_2$ with $0 \leq x < 0.5$ were prepared using reactions at opposite ends of the range of temperature and time ranges routinely used to make magnesium diboride. Despite the differences in the reactions, undoped samples appeared to be structurally very similar, and Williamson-Hall analyses of the broadening of x-ray diffraction peaks showed that both pure samples contained low levels of strain, representing a background level of crystalline imperfections. When Al was added, the derived lattice strains were not similar but varied according to the different synthesis heat treatment. A long reaction at 1200 °C produced samples with strain similar to that in the pure samples, indicating that the diffusion reactions were complete and that there was no systematic change in the background level of imperfections with $x$. However, a short reaction at 850 °C, which was intentionally stopped just before completion, produced samples with three times higher strain. The presence of $Mg_{17}Al_{12}$ and other Mg-Al intermetallics in these samples suggested that the higher strain is most likely due to Al composition inhomogeneity in the samples, which could be a result of the lower activity of Al than Mg. Since the superconducting transition was sharp for samples made by either reaction, the composition gradient of sample set *A* was thought to occur across all $(Mg,Al)B_2$ grains similarly and simultaneously.

The composition gradients skew the variation of superconducting properties as a function of Al concentration. For instance, two curves were found to describe $T_c(x)$, one for each sample set. By contrast, a single curve was observed when data were compared to the relative change of the unit cell volume. . Since the lattice constants, determined by x-ray diffraction, indicated the average Al content actually present within the superconducting grains, it was concluded that the latter comparison was the more accurate indication of intrinsic property changes. A similar contrast of data sets could be made for the superconducting gaps obtained by fitting heat capacity data. Yet, despite the skewing effect of the gradients, the critical temperature and heat capacity data were self-consistent for each sample set. It was argued that this makes it possible to misinterpret the response of the intrinsic superconducting properties with Al doping. The reaction that produced the composition gradients is similar to that which has been routinely used to make samples in previous Al doping studies, suggesting that previous conclusions should be cautiously evaluated.

## Acknowledgments

This work was supported by the U.S. Department of Energy, Division of Basic Energy Sciences under contract number DE-AC02-98CH1-886 and by BNL Laboratory-Directed Research and Development funds. We thank P. Canfield, R. Klie, D. Larbalestier, L. Lewis, M. Putti, R. Sabatini, D. Welch, R. Wilke, and Y. Zhu for stimulating discussions.